\def\/{\over}

\documentclass[preprint,aps,prd]{revtex4}
\newcommand{\bea}{\begin{eqnarray}}
\newcommand{\eea}{\end{eqnarray}}
\newcommand{\beq}{\begin{equation}}
\newcommand{\eeq}{\end{equation}}

\def\/{\over}

\begin{document}


\title{Spontaneous excitation of an accelerated hydrogen atom coupled with electromagnetic vacuum fluctuations
}

\author{ Zhiying Zhu${^2}$, Hongwei Yu$^{1,2,}$\footnote{Corresponding author} and Shizhuan Lu${^2}$ }
\affiliation{ $^1$CCAST(World Lab.), P. O. Box 8730, Beijing,
100080, P. R. China\\
$^2$Department of Physics and Institute of  Physics,\\
Hunan Normal University, Changsha, Hunan 410081, China
\footnote{Mailing address} }

\begin{abstract}
We consider a multilevel hydrogen atom in interaction with the
quantum electromagnetic field and separately calculate the
contributions of the vacuum fluctuation and radiation reaction to
the rate of change of the mean atomic energy of the atom for
uniform acceleration. It is found that the acceleration disturbs
the vacuum fluctuations in such a way that the delicate balance
between the contributions of vacuum fluctuation and radiation
reaction that exists for inertial atoms is broken, so that the
transitions to higher-lying states from ground state are possible
even in vacuum. In contrast to the case of an atom interacting
with a scalar field, the contributions of both electromagnetic
vacuum fluctuations and radiation reaction to the spontaneous
emission rate are affected by the acceleration, and furthermore
the contribution of the vacuum fluctuations contains a non-thermal
acceleration-dependent correction, which is possibly observable.
\end{abstract}

\maketitle

\baselineskip=16pt

\section{Introduction}
 Spontaneous emission is one of the most important features of
atoms and it may be attributed to vacuum fluctuations
\cite{Welton48,Compagno83}, or radiation reaction
\cite{Ackerhalt73}, or a combination of them \cite{Milonni75}. The
ambiguity in physical interpretation arises from different choices
of ordering of commuting operators of atom and field in a
Heisenberg picture approach to the problem. The ambiguity was
resloved, when Dalibard, Dupont-Roc and Cohen-Tannoudji showed
\cite{Dalibard82,Dalibard84} that there exists a symmetric
operator ordering that the distinct contributions of vacuum
fluctuations and radiation reaction to the rate of change of an
atomic observable are separately Hermitian and able to possess an
independent physical meaning.

Recently, Audretsch, M\"uller and Holzmann have generalized the
formalism of Ref.\cite{Dalibard84} to evaluate vacuum fluctuations
and radiation reaction contributions to the spontaneous excitation
rate \cite{Audretsch94} and radiative energy shifts
\cite{Audretsch95} of an accelerated two-level atom interacting
with a scalar field in a Minkowski vacuum. In particular, their
results show that when the atom is accelerated,  the delicate
balance between vacuum fluctuations and radiation reaction is
altered since the contribution of vacuum fluctuations to the rate
of change of the mean excitation energy is modified while that of
the radiation reaction remains the same. Thus transitions to
excited states for ground-state atoms become possible even in
vacuum. Based upon the formalism developed by Audretsch and
M\"uller\cite{Audretsch94},  the effects of modified vacuum
fluctuations and radiation reaction due to the presence of a
reflecting plane boundary upon the spontaneous excitation of both
an inertial and a uniformly accelerated atom interacting with a
quantized real massless scalar field have recently been discussed
 and similar conclusions are reached \cite{H. Yu}.

However, a two-level atom interacting with a scalar field is more
or less a toy model, and a more realistic system would be a
multi-level atom, a hydrogen atom, for instance, in interaction
with a quantized electromagnetic field. Such a system was recently
examined in terms of the radiative energy shifts of the
accelerated atom \cite{Passante97} using the method of
Ref.~\cite{Audretsch95}, where non-thermal corrections to the
energy shifts were found in addition to the usual thermal ones
associated with the temperature $T=a/2\pi$.  To make the spectrum
of research complete,  the atom's spontaneous emission rate in the
same realistic system will be considered in present paper
following the general method of Ref.\cite{Audretsch94}. We will
separately calculate the contribution of vacuum fluctuations and
the radiation reaction to the rate of variation of the atomic
energy of a hydrogen atom interacting with the quantum
electromagnetic field.  We will see that both the effects of
vacuum fluctuations and radiation reaction on the atom are changed
by the acceleration. This is in sharp contrast to the scalar field
case where the contribution of radiation reaction is not altered
by the acceleration.  A dramatic feature is that the contribution
of electromagnetic vacuum fluctuations to the spontaneous emission
rate contains an extra non-thermal term proportional to $a^2$, the
proper acceleration squared, in contrast to the scalar field case
where the effect of acceleration is purely thermal. Therefore the
equivalence between uniform acceleration and thermal fields is
lost when the scalar field is replaced by the electromagnetic
field as has been argued elsewhere in other different context
\cite{Boyer}.

\section{Interaction of a hydrogen atom and the
electromagnetic field}

Let us now briefly review the model introduced in
Ref.~\cite{Passante97}, where a linear interaction between a
hydrogen atom and the quantum electrmagnetic field is assumed. The
Hamiltonian that governs the time evolution of the atom with
respect to the proper time $\tau$ is written as
\begin{eqnarray}
H_A(\tau)=\sum_n\omega_n\sigma_{nn}(\tau)\;,
\end{eqnarray}
where $|n\rangle$ denotes a series of stationary atomic states
with energies $\omega_n$ and $\sigma_{nn}(\tau)=|n \rangle \langle
n|$. The units in which  $\hbar=c=1$ is adopted here and
hereafter. The free Hamiltonian of the quantum electromagnetic
field is
\begin{eqnarray}
H_F(\tau)=\sum_k \omega_{\vec{k}} a_{\vec{k}}^\dag a_{\vec{k
}}{dt\/d \tau}\;,
\end{eqnarray}
where $\vec{k}$ denotes the wave vector and polarization of the
field modes. We couple the hydrogen atom and the quantum
electromagnetic field in the multipolar coupling scheme
\cite{Passante97}
\begin{eqnarray}
H_I(\tau)=-e\textbf{ r}(\tau) \cdot
\textbf{E}(x(\tau))=-e\sum_{mn}\textbf{r}_{mn}\cdot
\textbf{E}(x(\tau))\;\sigma_{mn}(\tau)\;,
\end{eqnarray}
where $e$ is the electron electric charge, e$\textbf{r}$ the
atomic electric dipole moment,
$x(\tau)\leftrightarrow(t(\tau),\textbf{x}(\tau))$ the space-time
coordinates of the hydrogen atom. The Heisenberg  equations of
motion for the dynamical variables of the hydrogen atom and the
electromagnetic field can be derived from the Hamiltonian
$H=H_A+H_F+H_I$:
\begin{eqnarray}
{d\/d\tau}
\sigma_{mn}(\tau)=i(\omega_m-\omega_n)\sigma_{mn}(\tau)-ie\textbf{E}(x(\tau))\cdot
[\;\textbf{r}(\tau),\sigma_{mn}(\tau)\;]\;,
\end{eqnarray}
\begin{eqnarray}
{d\/d t}a_{\vec{k}}(t(\tau))&=&-i\omega_{\vec{k}}
a_{\vec{k}}(t(\tau))-ie\textbf{r}(\tau)\cdot
[\;\textbf{E}(x(\tau)),a_{\vec{k}}(t(\tau))\;]\;{d\tau\/dt}\;.
\end{eqnarray}
In the solutions of the equations of motion, we can separate the
``free" and ``source" parts,
\begin{equation}
\sigma_{mn}(\tau)=\sigma_{mn}^f(\tau)+\sigma_{mn}^s(\tau)\;, \quad
a_{\vec{k}}(t(\tau))=a_{\vec{k}}^f(t(\tau))+a_{\vec{k}}^s(t(\tau))\;,
\end{equation}
where
\begin{equation}
 \sigma_{mn}^f(\tau)=\sigma_{mn}^f(\tau_0)\;e^{i(\omega_m-\omega_n)(\tau-\tau_0)}\;,
\quad\sigma_{mn}^s(\tau)=-ie\int_{\tau_0}^\tau d\tau'\;
\textbf{E}^f(x(\tau'))\cdot[\;\textbf{r}^f(\tau'),\sigma_{mn}^f(\tau)\;]\;,
\end{equation}
\begin{equation}
a_{\vec{k}}^f(t(\tau))=a^f_{\vec{k}}(t(\tau_0))\;e^{-i\omega_{\vec{k}}[t(\tau)-t(\tau_0)]}\label{a1}\;,
\quad a_{\vec{k}}^s(t(\tau))=-ie\int_{\tau_0}^\tau d\tau'
\;\textbf{r}^f(\tau')\cdot[\;\textbf{E}^f(x(\tau')),a_{\vec{k}}^f(t(\tau))\;]\label{a2}\;.
\end{equation}

\section{The contributions of vacuum fluctuation and
radiation reaction  }

We now generalize the formalism of Ref.~\cite{Audretsch94} to the
model described in the proceeding section. We assume that the
initial state of the field is the vacuum $|0\rangle$, while the
atom is in the state $|b\rangle$. The equation of motion in the
interaction representation for an arbitrary atomic observable
$O(\tau)$, using symmetric ordering \cite{Dalibard82}, can be
separated in the vacuum fluctuations and the reaction field
contributions,
\begin{eqnarray}
{d O(\tau)\/d\tau}=\biggl({d O(\tau)\/
d\tau}\biggr)_{VF}+\biggl({d O(\tau)\/ d\tau}\biggr)_{RR}\;,
\end{eqnarray}
where
\begin{eqnarray}
\biggl({d O(\tau)\/ d\tau}\biggr)_{VF}=-{{ie}\/2}
\biggl(\textbf{E}^f(x(\tau))\cdot[\;\textbf{r}(\tau),O(\tau)\;]+
[\;\textbf{r}(\tau),O(\tau)\;]\cdot\textbf{E}^f(x(\tau))\biggr)\label{of}\;,
\end{eqnarray}
representing the contribution of the vacuum fluctuations and
\begin{eqnarray}
\biggl({d O(\tau)\/ d\tau}\biggr)_{RR}=-{{ie}\/2}
\biggl(\textbf{E}^s(x(\tau))\cdot[\;\textbf{r}(\tau),O(\tau)\;]+
[\;\textbf{r}(\tau),O(\tau)\;]\cdot\textbf{E}^s(x(\tau))\biggr)\label{os}\;,
\end{eqnarray}
denoting that of the radiation reaction.

Our purpose now is to identify the contributions of vacuum
fluctuations and radiation reaction in the evolution of the atom's
excitation energy, which is given by the expectation value of
$H_A$.  Separating $r_i(\tau)$ and $\sigma_{nn}(\tau)$ into their
free part and source part and taking the vacuum expectation value,
we can obtain, in a perturbation treatment up to order $e^2$,
\begin{eqnarray}
{\langle0|{dH_A(\tau)\/d
\tau}|0\rangle}_{VF}=-{e^2}\int_{\tau_0}^\tau
d\tau'\;C_{ij}^F(x(\tau),x(\tau'))[r_j^f(\tau')\;,
[\;r_i^f(\tau),\sum_n\omega_n\sigma_{nn}^f(\tau)]\;]\label{rhvf}\;,
\end{eqnarray}
\begin{eqnarray}
{\langle0|{dH_A(\tau)\/d
\tau}|0\rangle}_{RR}={e^2}\int_{\tau_0}^\tau
d\tau'\;\chi_{ij}^F(x(\tau),x(\tau'))\{r_j^f(\tau')\;,
[\;r_i^f(\tau),\sum_n\omega_n\sigma_{nn}^f(\tau)\;]\}\label{rhrr}\;.
\end{eqnarray}

The statistical functions $C_{ij}^F$ and $\chi_{ij}^F$ of the field
are defined as
\begin{eqnarray}
C_{ij}^F={1\/2}\langle0|\{E_i^f(x(\tau)),E_j^f(x(\tau'))\}|0\rangle\label{cf}\;,
\end{eqnarray}
\begin{eqnarray}
\chi_{ij}^F={1\/2}\langle0|[E_i^f(x(\tau)),E_j^f(x(\tau'))]|0\rangle\label{xf}\;,
\end{eqnarray}
$C_{ij}^F$ and $\chi_{ij}^F$ are the symmetric correlation function
and linear susceptibility of the field in the vacuum state.

We are interested in the evolution of expectation values of atomic
observables, so we take the expectation value of Eqs.(\ref{rhvf})
and (\ref{rhrr}) in the atom's state $|b\rangle$. Using the
Heisenberg equation of motion , we can replace the commutator
$[\;r_i^f(\tau),\sum_n\omega_n\sigma_{nn}^f(\tau)\;]$ with
$i{d\/d\tau}r_i^f(\tau)$, and obtain
\begin{eqnarray}
{\biggl\langle
{dH_A(\tau)\/d\tau}\biggr\rangle_{VF}}=2ie^2\int_{\tau_0}^\tau
d\tau' C_{ij}^F(x(\tau),x(\tau')){d\/d\tau}(\chi_{ij}^A)_b
(\tau,\tau')\label{hvf0}\;,
\end{eqnarray}
\begin{eqnarray}
{\biggl\langle
{dH_A(\tau)\/d\tau}\biggr\rangle_{RR}}=2ie^2\int_{\tau_0}^\tau
d\tau' \chi_{ij}^F(x(\tau),x(\tau')){d\/d\tau}(C_{ij}^A)_b
(\tau,\tau')\label{hrr0}\;,
\end{eqnarray}
where $|\rangle=|b,0\rangle$. Here symmetric correlation function
and linear susceptibility of the atom are defined analogously to
Eqs.(\ref{cf}) and (\ref{xf}) as
\begin{eqnarray}
(C_{ij}^A)_b(\tau,\tau')={1\/2}\langle
b|\{r_i^f(\tau),r_j^f(\tau')\}|b\rangle\;,
\end{eqnarray}\begin{eqnarray}
(\chi_{ij}^A)_b(\tau,\tau')={1\/2}\langle
b|[r_i^f(\tau),r_j^f(\tau')]|b\rangle\;.
\end{eqnarray}
They do not depend on the trajectory of the atom but characterize
only the atom itself. The explicit forms of the statistical
functions of the atom are given by
\begin{eqnarray}
(C_{ij}^A)_b(\tau,\tau')={1\/2}\sum_d[\langle
b|r_i(0)|d\rangle\langle d|r_j(0)|b\rangle
e^{i\omega_{bd}(\tau-\tau')}+\langle b|r_j(0)|d\rangle\langle
d|r_i(0)|b\rangle e^{-i\omega_{bd}(\tau-\tau')}]\label{ca}\;,
\end{eqnarray}\begin{eqnarray}
(\chi_{ij}^A)_b(\tau,\tau')={1\/2}\sum_d[\langle
b|r_i(0)|d\rangle\langle d|r_j(0)|b\rangle
e^{i\omega_{bd}(\tau-\tau')}-\langle b|r_j(0)|d\rangle\langle
d|r_i(0)|b\rangle e^{-i\omega_{bd}(\tau-\tau')}]\label{xa}\;,
\end{eqnarray}
where $\omega_{bd}=\omega_b-\omega_d$ and the sum extends over a
complete set of atomic states.

In order to get the statistical functions for the field, we will
use following  the two point function for the photon field in the
Feynman gauge
\begin{eqnarray}
\langle0|A_\mu(x)A_\nu(x^\prime)|0\rangle={\eta_{\mu\nu}\/{4\pi^2
[(t-t^\prime-i\varepsilon)^2-(x-x^\prime)^2-(y-y^\prime)^2-(z-z^\prime)^2]}}\;,
\end{eqnarray}
so, the field correlation function can be calculated as follows
\begin{eqnarray}
\langle0|E_i(x(\tau))E_j(x(\tau'))|0\rangle={1\/4\pi^2}(\partial
_0\partial_0^\prime\delta_{ij}-\partial_i\partial_j^\prime)
{1\/(x-x')^2+(y-y')^2+(z-z')^2-(t-t'-i\varepsilon)^2}\label{ee}\;,\nonumber\\
\end{eqnarray}
where $\varepsilon\rightarrow+0$ and $\partial^\prime$ denotes the
differentiation with respect to $x^\prime$.

\section{Uniformly accelerated atom}

Let us now apply the  formalism just developed to study the
spontaneous emission of the atom which is uniformly accelerated in
the $x$-direction. Specifically, the atom's trajectory is
described by
\begin{eqnarray}
t(\tau)={1\/a}\sinh a\tau\ ,\ \ \ x(\tau)={1\/a}\cosh a\tau\ ,\ \
\ y(\tau)=z(\tau)=0\label{tra}\;.
\end{eqnarray}
The field correlation function for the trajectory (\ref{tra}) can
be evaluated from its general form (\ref{ee}) in the frame of the
atom to get
\begin{eqnarray}
\langle0|E_i(x(\tau))E_j(x(\tau'))|0\rangle=\delta_{ij}
{1\/16\pi^2}{a^4\/\sinh^4[{a\/2}(\tau-\tau'-i\varepsilon)]}\label{ee2}\;.
\end{eqnarray}
From Eq.(\ref{ee2}), we obtain the symmetric correlation function
\begin{eqnarray}
C_{ij}^F(x(\tau),x(\tau'))=\delta_{ij}{a^4\/32\pi^2}
\biggl({1\/\sinh^4[{a\/2}(\tau-\tau'-i\varepsilon)]}
+{1\/\sinh^4[{a\/2}(\tau-\tau'+i\varepsilon)]}\biggr)
\end{eqnarray}
and the linear susceptibility

\begin{eqnarray}
\chi_{ij}^F(x(\tau),x(\tau'))=-i\delta_{ij}\,{1\/\pi\bigg(\cosh^3{a(\tau-\tau')\/2}
+5\cosh{a(\tau-\tau')\/2}\bigg)}\,\delta^{(3)}(\tau-\tau')\;,
\end{eqnarray}
where $\delta^{(3)}$ is the third derivative of the Dirac delta
function \footnote{Let us note that there is an error in Eq.~(3.4)
in Ref.~\cite{Passante97}}. With a substitution $u=\tau-\tau'$, we
get from (\ref{hrr0})
\begin{eqnarray}
{\biggl\langle{dH_A(\tau)\/d\tau}\biggr\rangle}_{RR}={ie^2\/\pi}\sum_d\omega_{bd}
|\langle
b|\textbf{r}(0)|d\rangle|^2\times\int^\infty_{-\infty}du\,{e^{i\omega_{bd}u}\/\cosh^3{a
u\/2} +5\cosh{a u\/2}}\,\delta^{(3)}(u)\;.
\end{eqnarray}
Here, we have extend the range of integration to infinity for
sufficiently long times $\tau-\tau_0$. After the evaluation of the
integral, we get
\begin{eqnarray}
{\biggl\langle{dH_A(\tau)\/d\tau}\biggr\rangle}_{RR}=
-{e^2\/6\pi}\biggl(\sum_{\omega_d<\omega_b} \omega_{bd}^4{|\langle
b|\textbf{r}(0)|d\rangle|}^2\biggl({a^2\/\omega_{bd}^2}+1\biggr)+\sum_{\omega_d>\omega_b}
\omega_{bd}^4{|\langle
b|\textbf{r}(0)|d\rangle|}^2\biggl({a^2\/\omega_{bd}^2}+1\biggr)\biggr)\label{hrr3}\;.\nonumber\\
\end{eqnarray}
The rate of change of the atomic energy is corrected by the
acceleration as compared to the inertial case ($a=0$) and it
always leads to a loss of energy of the atoms. In contrast to the
scalar field case \cite{Audretsch94}, where the uniform
acceleration does not change the contribution of the radiation
reaction, in our present case, there is an extra correction
proportional to $a^2$. The contribution of the vacuum fluctuation
to the rate of the atomic energy becomes now

\begin{eqnarray}
{\biggl\langle {d
H_A(\tau)\/d\tau}\biggr\rangle}_{VF}=-{{e^2a^4}\/32\pi^2}
\sum_d\omega_{bd} {|\langle b|\textbf{r}(0)|d\rangle|}^2\times
\int_{-\infty}^\infty
du\biggl({1\/\sinh^4[{a\/2}(u-i\varepsilon)]}+{1\/\sinh^4[{a\/2}(u+i\varepsilon)]}\biggr)
e^{i\omega_{bd}u}\label{hvf2}\;.\nonumber\\
\end{eqnarray}
One can calculate Eq.(\ref{hvf2}) by residues to get
\begin{eqnarray}
{\biggl\langle{dH_A(\tau)\over\,d\tau}\biggr\rangle}_{VF}&=&
-{e^2\/6\pi}\biggl(\;\sum_{\omega_d<\omega_b}
{\omega_{bd}^4\;|\langle
b|\textbf{r}(0)|d\rangle|}^2\;\biggl({a^2\/
\omega_{bd}^2}+1\biggr)\;\biggl(1+{2\/{e^{2\pi\omega_{bd}\/a}}-1}\biggr)
\nonumber\\&&-\sum_{\omega_d>\omega_b}{\omega_{bd}^4\;|\langle
b|\textbf{r}(0)|d\rangle|}^2\;\biggl({a^2\/
\omega_{bd}^2}+1\biggr)\;\biggl(1+{2\/{e^{2\pi|\omega_{bd}|\/a}}-1}\biggr)\;\biggr)\;.\label{hvf3}
\end{eqnarray}
This result reveals that vacuum fluctuations equally lead to
excitation of an accelerated ground-state atom and de-excitation
of an excited one and the probabilities of these two processes are
enhanced by the acceleration dependent correction terms as
compared to the inertial case. The most distinct feature in the
present case in contrast to that of a scalar field is that in
addition to a thermal term,  there exists an extra correction
proportional to $a^2$. The extra term is not in the form of a
thermal effect.  This is in sharp contrast with the scalar field
case \cite{Audretsch94} where the effect of acceleration upon the
contribution of vacuum fluctuations is only a ``thermal"
correction with the temperature $T=a/2\pi$. Therefore, one sees
that the equivalence between uniform acceleration and thermal
field is lost in the present case. For a typical transition
frequency in the Pfund Series of a hydrogen atom, $\omega\sim
10^{13}s^{-1}$ , the nonthermal correction would be comparable to
the thermal effect when the acceleration is approximately $\sim
\omega\sim 10^{23}cm/s^2 $. This acceleration, although extremely
large,  is at least of the same order of the acceleration
necessary to observe the thermal effect associated with uniform
acceleration in atomic systems \cite{Rosu}.

Finally, we add the contributions of vacuum fluctuations
(\ref{hrr3}) and radiation reaction (\ref{hvf3}) to obtain the
total rate of change of the atomic excitation energy:
\begin{eqnarray}
{\biggl\langle{dH_A(\tau)\over\,d\tau}\biggr\rangle}_{tot}&=&-{e^2\/3\pi}
\biggl(\sum_{\omega_d<\omega_b} {\omega_{bd}^4\;|\langle
b|\textbf{r}(0)|d\rangle|}^2\;\biggl({a^2\/
\omega_{bd}^2}+1\biggr)\;\biggl(1+{1\/{e^{2\pi\omega_{bd}\/a}}-1}\biggr)
\nonumber\\&&-\sum_{\omega_d>\omega_b}{\omega_{bd}^4\;|\langle
b|\textbf{r}(0)|d\rangle|}^2\;\biggl({a^2\/
\omega_{bd}^2}+1\biggr)\;{1\/{e^{2\pi|\omega_{bd}|\/a}}-1}\;\biggr).\label{tot2}
\end{eqnarray}
 For a uniformly accelerated atom in its ground state, only the second term contributes.
 This ensures the stability of ground-state atom at one hand and allows the transition to the excited states
 of accelerated ground-state atoms in the vacuum on the other.
\section{Conclusions}

In conclusion, assuming a linear coupling between a multi-level
atom and a quantum electromagnetic field,  we have studied the
contributions of vacuum fluctuations and radiation reaction to the
rate of change of the atomic energy of a uniformly accelerated
atom.  If the atom moves with constant acceleration, the perfect
balance between the contributions of vacuum fluctuations and
radiation reaction that ensures the stability of ground-state
atoms is disturbed, making spontaneous transition of ground-state
atoms to excited states possible. We find that the effects of
electromagnetic vacuum fluctuations on the spontaneous rate of an
accelerated atom is not purely thermal with the temperature
$T=a/2\pi$, and there exists a non-thermal correction, which might
be observable. This is in sharp contrast to the scalar field case
\cite{Audretsch94}. Another interesting feature worth noting, in
contrast to the scalar field case \cite{Audretsch94},  is that the
the acceleration also affects the contribution of radiation
reaction to the rate of change of atomic energy.

\begin{acknowledgments}
This work was supported in part  by the National Natural Science
Foundation of China  under Grants No.10375023 and No.10575035, the
Program for NCET (No.04-0784), the Key Project of Chinese Ministry
of Education (No.205110) and the Research fund of Hunan Provincial
Education Department (No. 04A030)
\end{acknowledgments}


\end{document}